# Chaos at Fifty

Adilson E. Motter and David K. Campbell

*In 1963 an MIT meteorologist revealed deterministic predictability to be an illusion and gave birth to a field that still thrives.*

*Adilson Motter is a professor of physics and astronomy at Northwestern University.*
*David Campbell is a professor of physics and of electrical and computer engineering at Boston University.*

In classical physics, one is taught that given the initial state of a system, all of its future states can be calculated. In the celebrated words of Pierre Simon Laplace, "An intelligence which could comprehend all the forces by which nature is animated and the respective situation of the beings who compose it—an intelligence sufficiently vast to submit these data to analysis . . . for it, nothing would be uncertain and the future, as the past, would be present to its eyes" [1]. Or, put another way, the clockwork universe holds true.

Herein lies the rub: *Exact* knowledge of a real-world initial state is never possible—the adviser can always demand a few more digits of experimental precision from the student, but the result will never be exact. Still, until the 19th century, the tacit assumption had always been that approximate knowledge of the initial state implies approximate knowledge of the final state. Given their success describing the motion of the planets, comets, and stars and the dynamics of countless other systems, physicists had little reason to assume otherwise.

Starting in the 19th century, however, and culminating with a 1963 paper by MIT meteorologist Edward Lorenz, pictured in figure 1a, a series of developments revealed that the notion of deterministic predictability, although appealingly intuitive, is in practice false for most systems. Small uncertainties in an initial state can indeed become large errors in a final one. Even simple systems for which all forces are known can behave unpredictably. Determinism, surprisingly enough, does not preclude chaos.

## A Gallery of Monsters

Chaos theory, as we know it today [2], took shape mostly during the last quarter of the 20th century. But researchers had experienced close encounters with the phenomenon as early as the late 1880s, beginning with Henri Poincaré's studies of the three-body problem in celestial mechanics. Poincaré observed that in such systems "it may happen that small differences in the initial conditions produce very great ones in the final phenomena. . . . Prediction becomes impossible" [3].

Dynamical systems like the three-body system studied by Poincaré are best described in phase space, in which dimensions correspond to the dynamical variables, such as position and momentum, that allow the system to be described by a set of first-order ordinary differential equations. The prevailing view had long been that, left alone, a conventional classical system will eventually settle toward either a steady state, described by a point in phase space; a periodic state, described by a closed loop; or a quasi-periodic state, which exhibits $n > 1$ incommensurable periodic modes and is described by an $n$-dimensional torus in phase space.

The three-body trajectories calculated by Poincaré fit into none of those categories. Rather, he observed that "each curve never intersects itself, but must fold upon itself in very complex fashion so as to intersect infinitely often each apex of the grid. One must be struck by the complexity of this shape, which I do not even attempt to illustrate," as paraphrased in English in ref. [4], p. 414.

What Poincaré refused to draw is now widely known as a homoclinic tangle, a canonical manifestation of chaos having fractal geometry. (An image of the tangle can be seen in figure 4 of the article by David Nolte, PHYSICS TODAY, April 2010, p. 33.)

Poincaré's results, independent findings by Jacques Hadamard, and experimental hints of chaos seen by their contemporaries were dismissed by many as pathologies or artifacts of noise or methodological shortcomings and were referred to as a "gallery of monsters" [4]. It would take nearly another century for chaos theory to gain a lasting foothold.

## A Serendipitous Discovery

In all likelihood, Lorenz was unfamiliar with Poincaré's work when he began his foray into meteorology in the mid 1900's (ref. [5], p. 133). With undergraduate and master's degrees in mathematics, Lorenz had served as a meteorologist in World War II before completing his doctoral studies in meteorology at MIT and joining the MIT faculty in 1955.

At the time, most meteorologists predicted weather using linear procedures, which were based on the premise that tomorrow's weather is a well-defined linear combination of features of today's weather. By contrast, an emerging school of dynamic meteorologists believed that weather could be more accurately predicted by simulating the fluid dynamical equations underlying atmospheric flows. Lorenz, who had just purchased his first computer, a Royal McBee LGP-30 with an internal memory of 4096 32-bit words, decided to compare the two approaches by pitting the linear procedures against a simplified 12-variable dynamical model. (Lorenz's computer, though a thousand times faster than his desk calculator, was still a million times slower than a current laptop.)

Lorenz searched for nonperiodic solutions, which he figured would pose the biggest challenge for the linear procedures, and eventually found them by imposing an external heating that varied with latitude and longitude—as does solar heating of the real atmosphere. Sure enough, the linear procedures yielded a far-from-perfect replication of the result.

Having found the nonperiodic solutions of his model interesting in their own right, Lorenz decided to examine them in more detail. He reproduced the data, this time printing the output variables after each day of simulated weather. To save space, he rounded them off to the third decimal place, even though the computer calculations were performed with higher precision. What followed is best appreciated in Lorenz's own words:

"At one point I decided to repeat some of the computations in order to examine what was happening in greater detail. I stopped the computer, typed in a line of numbers that it had printed out a while earlier, and set it running again. I went down the hall for a cup of coffee and returned after about an hour, during which time the computer had simulated about two months of weather. The numbers being printed were nothing like the old ones. I immediately suspected a weak vacuum tube or some other computer trouble, which was not uncommon, but before calling for service I decided to see just where the mistake had occurred, knowing that this could speed up the servicing process. Instead of a sudden break, I found that the new values at first repeated the old ones, but soon afterward differed by one and then several units in the last decimal place, and then began to differ in the next to the last place and then in the place before that. In fact, the differences more or less steadily doubled in size every four days or so, until all resemblance with the original output disappeared somewhere in the second month. This was enough to tell me what had happened: the numbers that I had typed in were not the exact original numbers, but were the rounded-off values that had appeared in the original printout. The initial round-off errors were the culprits; they were steadily amplifying until they dominated the solution." (ref. [5], p. 134)

## The Butterfly Effect

What Lorenz had observed with his model came to be known as sensitive dependence on initial conditions—a defining property of chaos. In phase space, the phenomenon has a distinct quantitative signature: The distance between any two nearby trajectories grows exponentially with time. Sensitive dependence is illustrated in figure 1b, one of Lorenz's own plots, which shows the gradual divergence of two time series calculated using identical equations but slightly different initial conditions. That trademark behavior gives chaotic systems the appearance of randomness. But as Lorenz himself noted, the appearances are deceiving: At any given time in a random system, one of two or more things can happen next, as one usually assumes for the throw of a die; in chaotic systems such as Lorenz's, outcomes are fully deterministic. (And strictly speaking, so are those of die throws—at least within a classical, macroscopic description.)

Lorenz realized that if the atmosphere were to behave like his model, forecasting the weather far in the future would be impossible. At a 1972 meeting of the American

Association for the Advancement of Science, in a talk titled "Predictability: Does the flap of a butterfly's wings in Brazil set off a tornado in Texas?" Lorenz used a butterfly as a metaphor for a tiny, seemingly inconsequential perturbation that could change the course of weather. The metaphor caught on, and sensitive dependence famously came to be dubbed the butterfly effect.

Given that computer simulations generally introduce round-off error at each time step—error that is amplified by chaos—one must ask whether Lorenz's solutions can possibly provide reliable information about real chaotic trajectories. As it happens, they can, because of a property now known as shadowing: Although for any given initial condition the numerical trajectory diverges from the exact one, there always exists a nearby initial condition whose exact trajectory is approximated by the numerical one for a prespecified stretch of time. In the end, it is as if one had started from a different initial condition and calculated the trajectory exactly—a crucial result, given that numerical calculations came to be widely used in the study of chaotic systems. For example, the trajectories Lorenz calculated using the truncated variables were, in fact, just as representative of his model's behavior as the original (as well as the exact) trajectories.

Lorenz first presented the results from his 12-variable model at a 1960 symposium held in Tokyo. At that meeting, he only briefly mentioned the unexpected effect of round-off errors; he believed those results belonged in a different paper. In retrospect, he was in little danger of being scooped—apparently, most of his contemporaries failed to recognize the broad significance of his findings. (Meanwhile, the work of other pioneers of chaos often went unappreciated; see ref. [6] for Yoshisuke Ueda's description of his frustration at the lack of appreciation of his 1961 analog computer observations of the "randomly transitional phenomenon," later recognized as chaos.)

## The Lorenz Attractor

Lorenz published his serendipitous discovery in a March 1963 paper titled "Deterministic nonperiodic flow" [7]. He had spent a significant part of his time since the Tokyo meeting looking for the simplest possible model exhibiting sensitive dependence on initial conditions, and he eventually arrived at a three-variable system of nonlinear ordinary differential equations now known as the Lorenz equations (see Box 1).

Like Poincaré's three-body system, the Lorenz equations yield phase-space trajectories that never retrace themselves and that don't trace out surfaces of integer dimension. Rather, typical trajectories tend to converge to, and then orbit along, a bounded structure of non-integer, fractal dimension known as a chaotic attractor. (See figure 2.)

Perhaps the most studied objects in chaos theory, chaotic attractors tend to emerge when a dissipative system is regularly forced to compensate for the loss of energy—as when a child in a swing kicks his or her legs to keep the motion going. In the case of the Lorenz system, forcing is by way of heating, and dissipation is due to the viscosity of the fluid.

A chaotic attractor is the example *par excellence* of a chaotic set. A chaotic set has uncountably many chaotic trajectories; on such a set, any point that lies in the neighborhood of a given point will also, with probability one, give rise to a chaotic trajectory. Yet no matter the proximity of those two points, in the region between them will lie points of infinitely many periodic orbits. In mathematical parlance, the periodic orbits constitute a countable, zero-measure, but dense set of points embedded in the chaotic set, analogous to the rational numbers embedded in the set of real numbers. Not only will trajectories that lie on the attractor behave chaotically, any point lying within the attractor's basin of attraction will also give rise to chaotic trajectories that converge to the attractor.

If chaotic sets such as the Lorenz attractor are embedded with infinitely many periodic orbits, why doesn't one ever see those orbits in practice? The answer, and the key feature underlying chaos, is that the periodic orbits are unstable; they cause nearby orbits to diverge, just as the trajectories of a simple pendulum diverge in the neighborhood of the unstable "up" position. But whereas the pendulum trajectories diverge at one point, periodic orbits embedded in the chaotic set cause trajectories to diverge at every point. That skeleton of unstable periodic orbits is what leads to the irregular, chaotic dynamics seen in Lorenz's model and other chaotic systems. Lorenz appears to have grasped that essential feature of chaos early on; he recognized not only that nonperiodicity implies sensitive dependence but that sensitive dependence is the root cause of nonperiodicity.

One might have expected Lorenz's seminal publication—a model of clarity and concision—to have attracted immediate attention. It did not. Twelve years after its publication, the paper had accumulated fewer than 20 citations. The turning point was when mathematicians and physicists learned of the work, largely through Tien-Yien Li and James Yorke's 1975 paper, "Period three implies chaos" [8], which established the name of the field, albeit with a slightly more restrictive meaning than it has today. By the late 1980s, not only had research on chaos skyrocketed, as evidenced by the thousands of scientific publications on the topic, it was already being widely popularized among nonscientists [9].

## Fractals, Folding, and *Mille Feuille*

Chaotic attractors are generally fractals. The relationship between the chaotic and fractal aspects can be understood by considering the trajectories of a blob of points in the phase space near a chaotic attractor. The chaotic dynamics on the attractor stretches the blob in some directions and contracts it in others, thus forming a thin filament. But because the trajectories are bounded, the filament must eventually fold on itself. When that sequence of stretching and folding is repeated indefinitely—analogous to a baker kneading dough or preparing *mille feuille* pastry—it gives rise to a fractal set, for which the distance between two typical points from the original blob, measured along the resulting attractor, is infinite. An attractor's geometry can be quantitatively related to its dynamical properties: The (fractal) dimension can be extracted, for example, from the rate at which nearby trajectories diverge in phase space or from the time series of a single variable [10]. Physically, the fractal dimension represents the effective number of degrees of freedom a system has once it has settled on the attractor. Although Lorenz could not resolve it with his numerics, his attractor has a fractal dimension of roughly 2.06.

Figure 3a shows the asymptotic behavior of the phase-space trajectories of another chaotic attractor, that of the periodically driven, damped pendulum. The fractal nature of the attractor can be seen by zooming in on a small portion of phase space: On magnification, the attractor appears statistically self-similar. Given the intimate relationship between the chaotic and fractal natures of such attractors, it was more than a coincidence that the study of fractals reached its maturity in the 1970s, just as chaos was becoming widely known.

Chaos can also find its way into dissipative systems whose attractors are not chaotic, as is the case for a periodically driven pendulum that's very strongly damped. The phase space of such a system, depicted in figure 3b, contains two periodic attractors, corresponding to clockwise and counterclockwise rotations of the pendulum. Typical trajectories converge to one of the two attractors with probability one, with each attractor having its own distinct basin of attraction, as illustrated in the figure. However, embedded at the boundary between those basins of attraction there is a zero-measure, fractal chaotic set—a repeller—that transiently influences the evolution of nearby trajectories. A similar phenomenon may occur in systems whose trajectories are unbounded, as in chaotic scattering processes.

## Hamiltonian Chaos

As foreshadowed by Poincaré, chaos can also appear in conservative systems, such as those described by Hamiltonians. Unlike in dissipative systems, where a high-dimensional basin of attraction may converge to a lower-dimensional attractor, in conservative systems trajectories necessarily conserve volume in phase space.

To understand how chaos arises in a conservative system, consider a Hamiltonian system—a chain of frictionless harmonic oscillators, say—with $n$ degrees of freedom. The system is integrable, and hence nonchaotic, if it has $n$ independent integrals of motion—that is, if it is described by $n$ conserved quantities such as energy and momenta. If the trajectories are bounded, the system's motion will be constrained to surfaces that are topologically equivalent to $n$-dimensional tori; each dimension of a torus is associated with a periodic mode of the system. A generic perturbation of the Hamiltonian will destroy resonant tori, for which the various periodic modes have frequency ratios that are easily approximated by rational numbers. Some of the corresponding orbits gain access to $2n$–dimensional regions of the phase space and become chaotic; others form new families of smaller-scale tori. The resonant tori in the new families are destroyed by the same mechanism, and so on. The Kolmogorov-Arnold-Moser theorem guarantees that non-resonant tori survive the perturbation, but the fraction of tori, and hence orbits, that fall into that category decreases with the strength of the perturbation. The end result is that the phase space of a generic Hamiltonian system contains coexisting regular and chaotic regions, which extend to arbitrarily small scales (see figure 4).

Beautiful manifestations of Hamiltonian chaos are visible in the asteroid belt and in the rings of Saturn, where unpopulated gaps correspond to chaotic trajectories that were unconfined to the nearly circular, ring-like orbits (see figure 5a).

## Bifurcations and Universality

Dynamical systems commonly exhibit bifurcations—sudden changes in behavior as a parameter of the system is varied, such as the sudden onset of convection rolls in a fluid heated from below once the temperature gradient exceeds some threshold. A decisive moment in the development of chaos theory came in the late 1970s, when high-precision experimental methods in fluids (see the article by Harry Swinney and Jerry Gollub, PHYSICS TODAY, August 1978, p. 41) and novel numerical and statistical-physics techniques allowed researchers to explore in quantitative detail how chaos can arise through various sequences of bifurcations.

Mitchell Feigenbaum showed in 1978 that for a wide class of mathematical and experimental systems, one such sequence of bifurcations—the so-called period-doubling route to chaos—occurs the same way, at the same normalized values of the bifurcation parameter. That particular form of universality was subsequently demonstrated in low-temperature convection experiments by Albert Libchaber and Jean Maurer, a development that sparked an explosion of interest in chaos and earned Feigenbaum and Libchaber the 1986 Wolf Prize in Physics [11]. Since then, theoretical and experimental studies have confirmed the universality of period doubling in a variety of systems, including in the Lorenz equations themselves.

## What have we Learned?

Chaos sets itself apart from other great revolutions in the physical sciences. In contrast to, say, relativity or quantum mechanics, chaos is not a theory of any particular physical phenomenon. Rather, it is a paradigm shift of all science, which provides a collection of concepts and methods to analyze a novel behavior that can arise in a wide range of disciplines. Those traits partly explain the indifference with which the initial hints of the phenomenon were greeted: Early encounters with chaos took place in disparate disciplines—celestial mechanics, mathematics, and engineering—whose practitioners were not aware of each other's findings. Also, chaos generally defies direct analytic approaches. Only when advances in interactive computation made experimental mathematics [12] a reality could one pursue the insights of Poincaré and the other pioneers.

The basics of chaos have been incorporated into physics and applied mathematics curricula, but strong interest remains in understanding specific manifestations of the phenomenon in fields ranging from applied physics and engineering to physiology, computer science, and finance [13]. For instance, a recent study reexamining a long-standing debate suggests that a healthy human heartbeat is chaotic due to coupling with breathing [14], much as a star–planet system can become chaotic in the presence of a second planet.

Another body of research has established that, despite sensitive dependence on initial conditions, coupled chaotic systems can synchronize on a shared chaotic trajectory [15], a phenomenon with many applications in networked systems. (See the article by Adilson E. Motter and Réka Albert, PHYSICS TODAY, April 2012, p. 43.) Other work has established relations between chaos and the so-called *P* versus *NP* problem in computer science. In particular, it has been shown that constraint-optimization problems can be described in terms of dynamical systems that become transiently chaotic as optimization hardness increases [16].

Perhaps no field of research can benefit as much from the study of chaos as fluid dynamics. Even in flows governed by periodic velocity fields, microscale fluid elements often move chaotically. A classic example is the transient chaotic behavior of a flow past an obstacle, a behavior that has been proposed to explain how competing plankton species coexist in certain island locales. In a well-mixed environment, all but a handful of species would go extinct. But in the flows that emerge in an island's wake, the various species can inhabit different fractal-like flow structures of high surface-to-volume ratio that may intertwine but do not mix [17]. (See figure 5b.) Similarly, stretching, folding, and the exponential separation of nearby points—all hallmarks of chaos—are observed in Lagrangian coherent structures, which are of interest, for example, to forecasting contaminant transport in the ocean and atmosphere. (See the article by Thomas Peacock and George Haller, PHYSICS TODAY, February 2013, p. 41.)

Although low-dimensional chaos does not speak directly to turbulence, spatiotemporal chaos is observed in flows at high Reynolds numbers. Fittingly, Lorenz made the connection between chaos and turbulence at the very beginning—his first choice for the title of his seminal 1963 paper was, in fact, "Deterministic turbulence," which he abandoned only at the urging of the editor.

Numerous fundamental problems in chaos remain at least partially unsettled. They range from the implications of chaos in quantum and relativistic systems to the connection between chaos, coarse graining in phase space, and statistical mechanics. Another fundamental activity concerns model building. For example, the irregular polarity reversals observed at astronomical time scales in Earth's magnetic field have recently been described with a deterministic chaotic model not unlike the three-equation model that begat the field a half-century ago (see figure 5c).

The Lorenz attractor has turned out to be representative of the asymptotic dynamics of many systems, and Lorenz's signature contribution has reverberated both broadly and deeply. As summarized in the citation of his 1991 Kyoto Prize, "He made his boldest scientific achievement in discovering 'deterministic chaos,' a principle which has profoundly influenced a wide range of basic sciences and brought about one of the most dramatic changes in mankind's view of nature since Sir Isaac Newton."

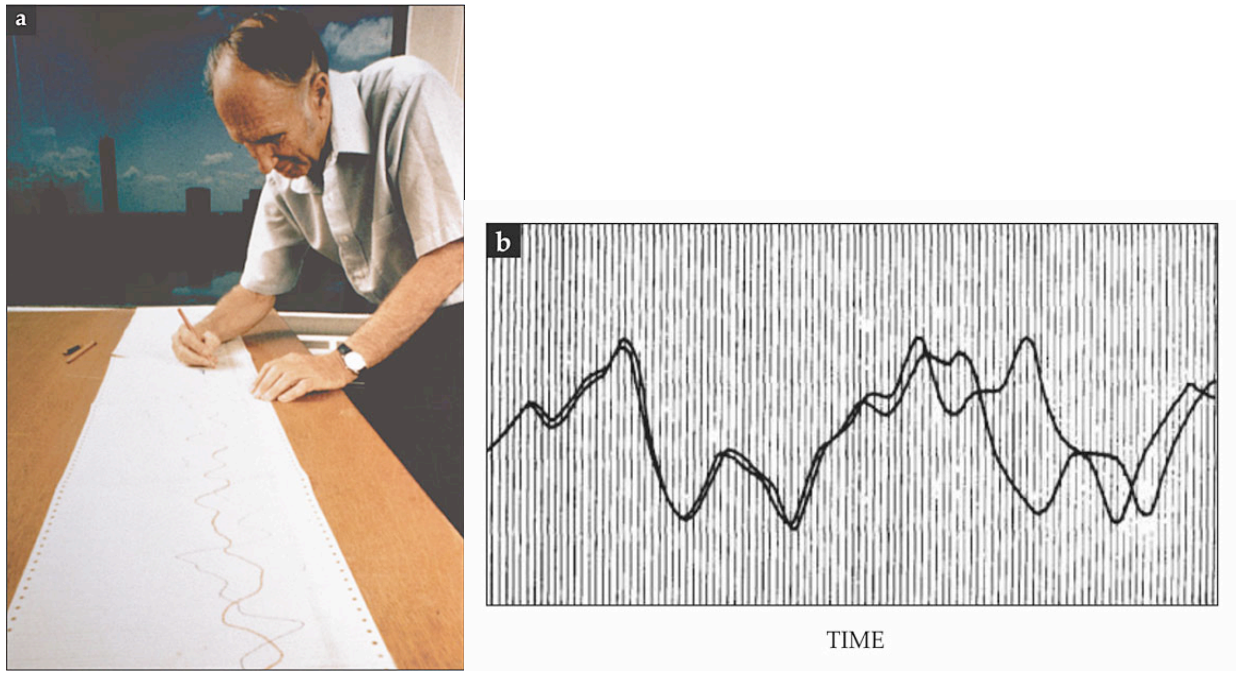

**Figure 1**. Edward Lorenz and the butterfly effect. **(a)** Lorenz, studying a computer-generated time series. (Photo courtesy of the Inamori Foundation.) **(b)** A close-up of Lorenz's original printout from his discovery of the butterfly effect shows two time series generated with the same equations but with slightly different initial conditions. The series diverge exponentially with time due to sensitive dependence on initial conditions. (Adapted from ref. [9].)

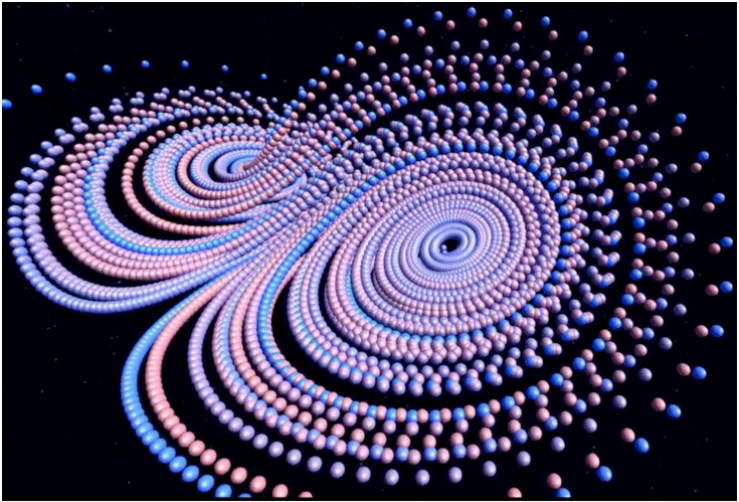

**Figure 2**. The Lorenz attractor, as revealed by the never-repeating trajectory of a single chaotic orbit. The spheres shown here represent iterations of the so-called Lorenz equations, calculated using the original parameters in Edward Lorenz's seminal work. (Spheres are colored according to the iteration count.) From certain angles, the two lobes of the attractor resemble a butterfly, a coincidence that helped earn sensitive dependence on initial conditions its nickname—the butterfly effect. For an animated visualization of the attractor, see http://www.youtube.com/watch?v=iu4RdmBVdps. (Image courtesy of Stefan Ganev.)

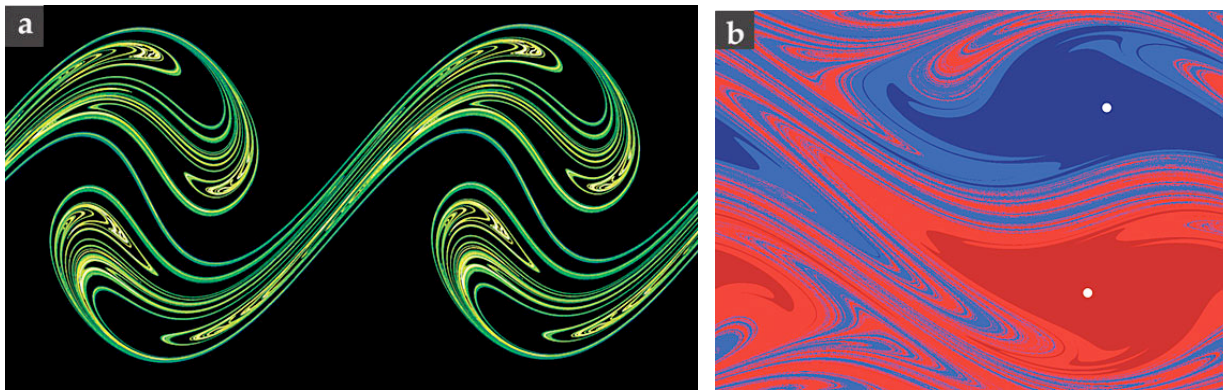

**Figure 3.** Chaos in dissipative systems. **(a)** The phase-space trajectories of a periodically driven, damped pendulum converge to a chaotic attractor, plotted here at integer multiples of the driving period; stretching and folding of volumes in phase space gives the attractor its fractal structure. **(b)** For a sufficiently dissipative pendulum, the phase space contains two nonchaotic, periodic attractors, indicated here with white dots on the plane of initial conditions. Nevertheless, the phase space contains a chaotic set at the boundary between the attractors' respective basins of attraction, indicated in red and blue. In both panels, the *x* and *y* dimensions are position (angle) and angular momentum, respectively. For visualization purposes, the horizontal axis in panel (a) is plotted over more than one period of the angle. (Adapted from T. Tel, M. Gruiz, *Chaotic Dynamics: An Introduction Based on Classical Mechanics*, K. Kulacsy, trans., Cambridge U. Press, New York, 2006.)

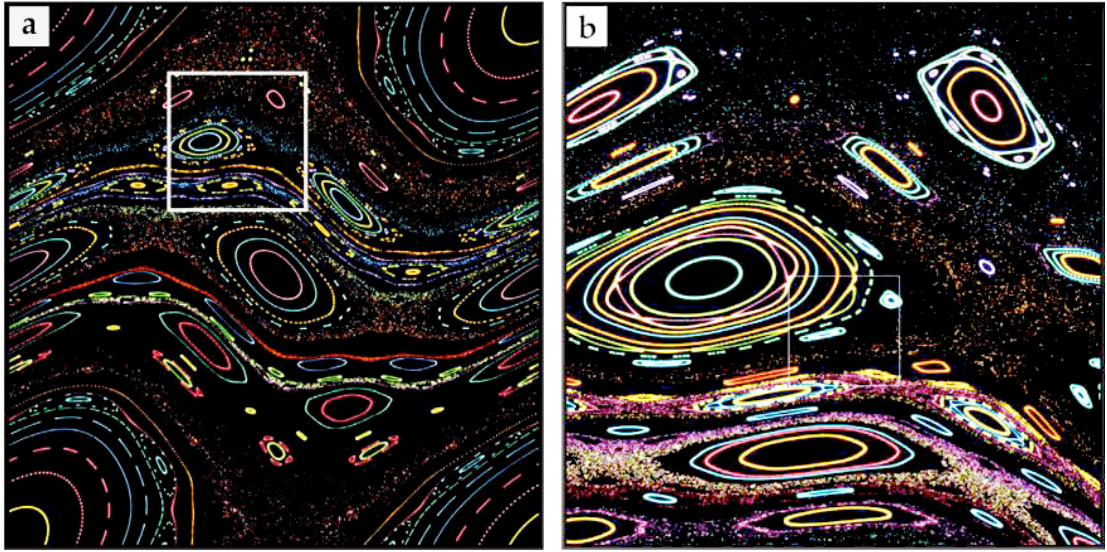

**Figure 4.** Chaos in conservative systems. **(a)** A stroboscopic map shows the phase-space trajectories of a periodically kicked rotor. The map displays periodic and quasi-periodic regions, which correspond to the looped trajectories in the image, and chaotic regions, which correspond to the scattered trajectories. Each trajectory is indicated by points of a single color. **(b)** Magnification of the boxed region in the phase space illustrates the approximately self-similar nature of the phase space. In both panels, the *x* and *y* dimensions are position and angular momentum, respectively. (Adapted from D. K. Campbell, in *From Cardinals to Chaos: Reflections on the Life and Legacy of Stanislaw Ulam*, N. G. Cooper, ed., Cambridge U. Press, New York, 1989.)

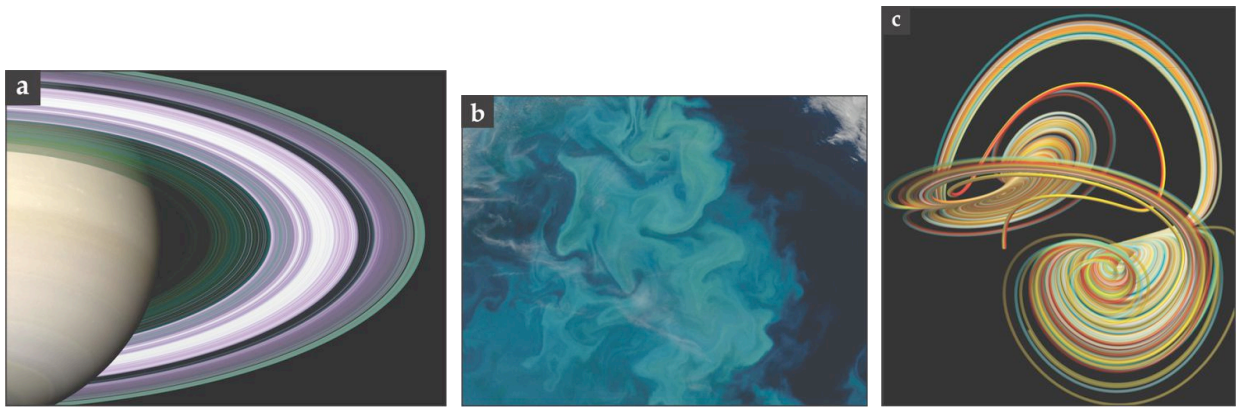

**Figure 5.** Chaos manifests itself in a diverse range of natural settings, including **(a)** the rings of Saturn, where unpopulated gaps correspond to chaotic orbits, as predicted by the Kolmogorov-Arnold-Moser theory (courtesy of NASA/ Cassini mission); **(b)** phytoplankton blooms—seen here in a satellite image of the Barents Sea—which form fractal-like structures due to chaotic advection (adapted from NASA/Ocean Color Web); and **(c)** Earth's geomagnetic field, which, on astronomical time scales, reverses its poles at irregular, chaotic intervals—a behavior captured by this simulated chaotic attractor (courtesy of Christophe Gissinger).

## BOX 1: The Lorenz equations

The three-equation model used by Edward Lorenz to demonstrate chaos derives from a truncated Fourier series expansion of the partial differential equations describing a thin, horizontal layer of fluid heated from below and cooled from above. Lorenz proposed the equations as a crude model of the motion of a region of the atmosphere driven by solar heating of Earth. In standard notation, the equations are:

$$dX/dt = \sigma(-X + Y),$$
$$dY/dt = rX - Y - XZ,$$
$$dZ/dt = -bZ + XY,$$

where $X$ represents the intensity of the convective motion, $Y$ is proportional to the temperature difference between the ascending and descending convective currents, and $Z$ indicates the deviation of the vertical temperature profile from linearity. The parameters $b$ and $\sigma$ capture particulars of the flow geometry and rheology, and $r$, the Rayleigh number, determines the relative importance of conductive and convective heat transfer. Lorenz fixed $b$ and $\sigma$ at 8/3 and 10, respectively, leaving only $r$ to vary. For small $r$, the system has a stable fixed point at $X = Y = Z = 0$, corresponding to no convection. At $r = 1$, two symmetrical fixed points, representing two steady convective states, emerge. For $r \gtrsim 24.74$, the convective states lose stability, and at $r = 28$, the system exhibits nonperiodic trajectories like the one shown in figure 2 of the text. Such trajectories forever orbit along a bounded region of the three-dimensional space known as a chaotic attractor and never intersect themselves—otherwise they would be periodic. For larger values of $r$, the Lorenz equations exhibit a remarkable array of different behaviors, which are carefully cataloged in ref. [18].